\documentclass[a4paper,twoside]{article}

\usepackage{epsfig}
\usepackage{subfigure}
\usepackage{calc}
\usepackage{amssymb}
\usepackage{amstext}
\usepackage{amsmath}
\usepackage{amsthm}
\usepackage{multicol}
\usepackage{pslatex}
\usepackage{apalike}
\usepackage{SCITEPRESS}    

\usepackage[hyphens]{url}
\usepackage[hidelinks]{hyperref}
\hypersetup{breaklinks=true}

\renewcommand{\thesubsubsection}{\thesection.\Alph{subsubsection}}

\usepackage{array}
\usepackage[T1]{fontenc}

\newcolumntype{P}[1]{>{\raggedright\arraybackslash}p{#1}}
\newcolumntype{U}{S[table-format=1.3,table-column-width=4em]}
\usepackage{tabularx}
\usepackage{multirow}

\usepackage{listings}

\makeatletter
\def\thickhline{%
  \noalign{\ifnum0=`}\fi\hrule \@height \thickarrayrulewidth \futurelet
   \reserved@a\@xthickhline}
\def\@xthickhline{\ifx\reserved@a\thickhline
               \vskip\doublerulesep
               \vskip-\thickarrayrulewidth
             \fi
      \ifnum0=`{\fi}}
\newcolumntype{!}{@{\vrule width 1pt}}
\makeatother

\newlength{\thickarrayrulewidth}
\begin{document}

\title{Responding to Living-Off-the-Land Tactics using \\Just-in-Time Memory Forensics (JIT-MF) for Android}

\author{\authorname{Jennifer Bellizzi\sup{1}\orcidAuthor{0000-0003-1754-9473}, Mark Vella\sup{1}\orcidAuthor{0000-0002-6483-9054}, Christian Colombo\sup{1}\orcidAuthor{0000-0002-2844-5728} and Julio Hernandez-Castro\sup{2}\orcidAuthor{0000-0002-6432-5328}}
\affiliation{\sup{1}Department of Computer Science, University of Malta, Msida, Malta}
\affiliation{\sup{2}School of Computing, Cornwallis South, University of Kent, Canterbury, UK}
\email{\{jennifer.bellizzi, mark.vella, christian.colombo\}@um.edu.mt, jch27@kent.ac.uk}
}


\abstract{
Digital investigations of stealthy attacks on Android devices pose particular challenges to incident responders. Whereas consequential late detection demands accurate and comprehensive forensic timelines to reconstruct all malicious activities, reduced forensic footprints with minimal malware involvement, such as when Living-Off-the-Land (LOtL) tactics are adopted, leave investigators little evidence to work with. Volatile memory forensics can be an effective approach since app execution of any form is always bound to leave a trail of evidence in memory, even if perhaps ephemeral. Just-in-Time Memory Forensics (JIT-MF) is a recently proposed technique that describes a framework to process memory forensics on existing stock Android devices, without compromising their security by requiring them to be rooted. Within this framework, JIT-MF drivers are designed to promptly dump in-memory evidence related to app usage or misuse. In this work, we primarily introduce a conceptualized presentation of JIT-MF drivers. Subsequently, through a series of case studies involving the hijacking of widely-used messaging apps, we show that when the target apps are forensically enhanced with JIT-MF drivers, investigators can generate richer forensic timelines to support their investigation, which are on average 26\% closer to ground truth.}

\onecolumn \maketitle \normalsize \setcounter{footnote}{0} \vfill

\section{\uppercase{Introduction}}\label{sec:intro}
The use of process memory forensics is increasingly becoming a necessity when investigating advanced stealthy cyberattack incidents targeting smartphones \cite{case2020hooktracer,ali2020app,bhatia2018tipped,taubmann2018droidkex}. This is the reality incident responders face amidst the limitations and barriers of the more forensically sound, but alas limited, state-of-the-art mobile forensics.
In this work, our primary concern is Android malware that exhibits long-term stealth by evading early detection mechanisms, which eventually is only detected through its consequences. Specifically, we focus on attacks that hijack the messaging functionality of Android devices to hide compromising communication of a criminal nature behind victim devices or else spy on them through unlawful interception. In the eventuality that device owners notice suspicious activity and hand their device over for further investigation, incident responders would need to investigate --- utilizing a forensic timeline \cite{plaso} --- all the activities undertaken by threat actors through the deployed malware.

Android accessibility trojans are a case in point \cite{defensorid,androidrat}. This attack vector has been shown to enable stealthy Living-Off-the-Land (LOtL) tactics \cite{lotl}, where key attack steps get delegated to benign apps, possibly only requiring the use of malware during an initial setup phase. In similar scenarios, trying to establish a forensic timeline solely from sources found on non-volatile flash memory, whether on-chip or removable, can prove futile even after many barriers to forensic evidence collection are overcome.
The root cause is that the forensic artefacts constituting attack evidence would have been erased from storage or never even created to begin with. Volatile memory can be an effective forensic source in such circumstances. No matter how stealthy an attack can be, its execution through malware or benign victim apps has to occur in memory \cite{case2017memory}. Therefore, any resulting evidence has to be present in process memory, even if just briefly. While full-device static dumps of volatile memory typically require an unlocked bootloader and customized firmware, dynamic dumps of process memory do not necessarily face similar restrictions \cite{bellizzi2021real}. We show that this is the case when deployed using a mix of static and dynamic app instrumentation, at least when not involving system apps and services.

Just-in-time Memory Forensics (JIT-MF) is a framework for live process memory forensics, with an initial study describing how to formulate and implement JIT-MF drivers \cite{bellizzi2021real}. These drivers are responsible for establishing the points in time when memory dumps should be triggered and the heap/native memory areas/objects to be included. While in this paper we implement drivers in the context of message hijacking, these are meant to be configurable to cater for other investigative scenarios unrelated to message hijacking.

\begin{figure}
\centering
\includegraphics[page=1,trim = 12mm 40mm 35mm 15mm, clip, width=1\linewidth]{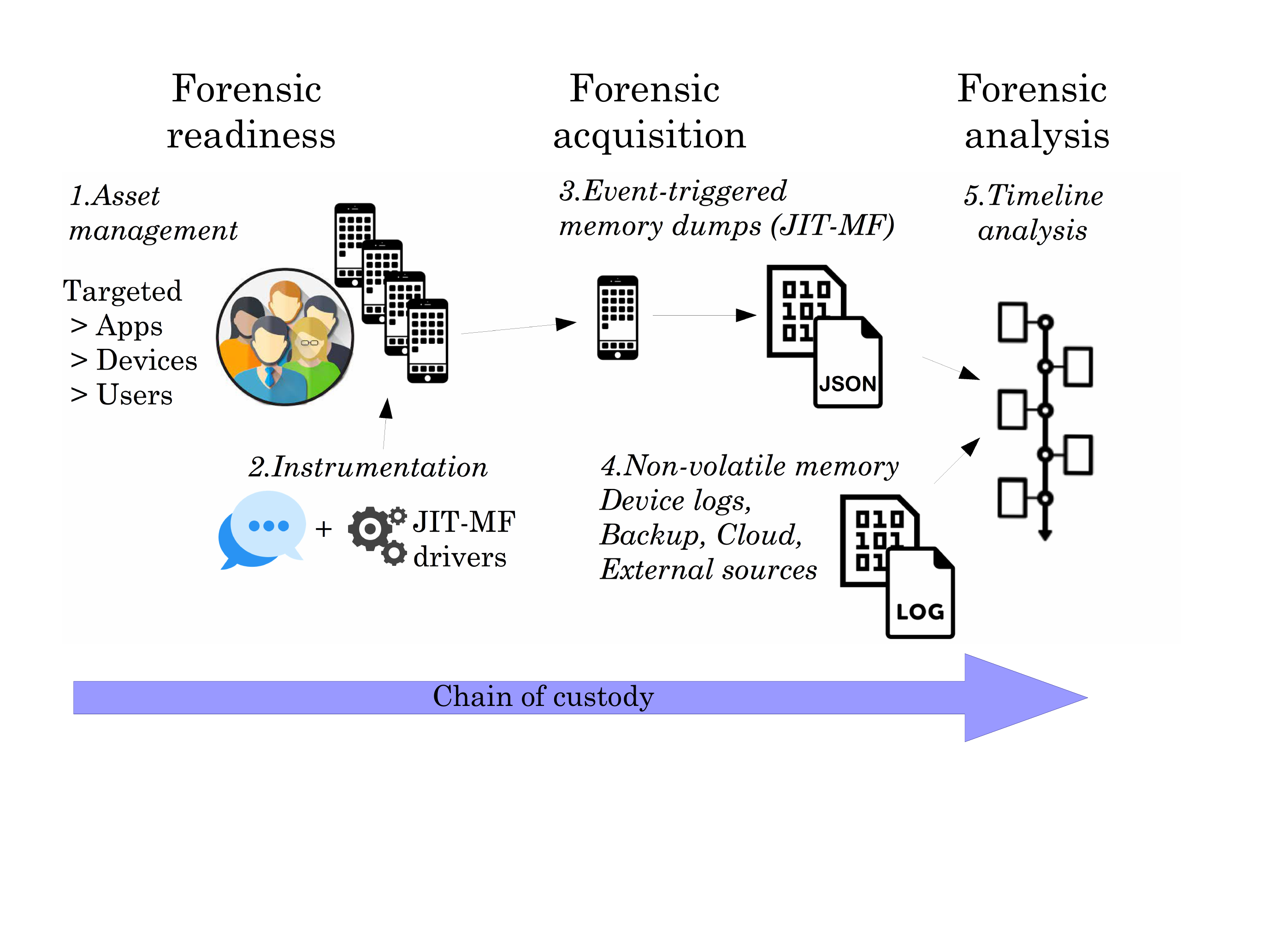}
\caption{The JIT-MF workflow.}
\label{fig:jitmf}
\end{figure}
Figure \ref{fig:jitmf} shows the complete JIT-MF workflow. Starting with a forensic readiness stage \cite{irbook}, targeted users along with their devices and apps are identified during an asset management exercise (step 1). These users can be high-profile employees of government agencies or even private citizens whose devices may be the target of resourceful attackers for various reasons. Once a risk assessment is carried out, those apps that pose a particular risk, say messaging apps, are instrumented with JIT-MF drivers (step 2). With JIT-MF, the forensic acquisition of memory dumps is triggered by specific app events identified as trigger points by JIT-MF drivers (step 3). These memory dumps contain either raw binaries from memory segments or else readily-carved and parsed objects along with tagged metadata, e.g.\ in JSON format. These two acquisition methods are referred to as offline and online respectively, depending on how object carving and parsing are carried out. 

Once suspicious activity is noticed, with alerts possibly raised by the device owners themselves or by incident responders during routine checks, the JIT-MF dumps are merged with other forensic sources to produce a forensic timeline (steps 4 and 5). JIT-MF's scope is that of producing a richer forensic timeline, aiming for a comprehensive reconstruction of app activity, supporting investigators to establish the full picture of the incident. Techniques, such as comparisons with a baseline of known normal device/app usage or cross-checking with cyber threat intelligence feeds, can then be used to identify attack-related, non-consented app usage. The entire workflow steps should be recorded, accompanied with full metadata, hashes, and digital signatures, thus keeping a complete chain of custody \cite{cosic2010framework}. In this manner, all the acquired evidence remains admissible to a court of law if need be.

JIT-MF drivers can be deployed on stock non-rooted devices without requiring any firmware alterations and are compatible with encrypted devices. By avoiding an approach based entirely on forensically enhancing the Android kernel or individual app codebases, JIT-MF can work with what is already deployed in the field. This approach avoids making any form of imposition on future releases of Android and/or apps developed for the platform. While not clashing with device access controls, such as screen and bootloader locks, JIT-MF does require the device owner's collaboration. This assumption's realism rests on the fact that device owners are the potential victims rather than perpetrators.

The key proposition of this paper is that through JIT-MF tools and their drivers, timely captured volatile memory can be used as an additional forensic source to help reconstruct incident scenarios in a more comprehensive manner, better supporting incident investigations that involve stealthy, long-running cyberattacks targeting Android smartphones, and their owners alike. We make the following contributions:

\begin{itemize}
    \item Provide a conceptual, generic description of JIT-MF drivers, along with a methodology for their implementations.
    \item Six JIT-MF drivers for popular messaging apps, covering both SMS and instant messaging, supporting investigators to determine any malicious/non-consented activity.
    \item Experimentation involving stealthy messaging hijack case studies, demonstrating how using JIT-MF drivers results in richer forensic timelines, as compared to solely relying on forensic sources used by state-of-the-art mobile forensics tools.
\end{itemize}

\vspace{-0.3in}
\section{Background}\label{sec:background}
Given the nature of apps that are typically installed on smartphones, ranging from messaging, voice/video calls, to the camera, navigation, calendar, and social networking, just to name a few, mobile devices are nowadays both a rich source of evidence as well as a primary target for cyberattacks \cite{scrivens2017android}.

\subsection{Android forensics}\label{sec:androidforensics}
\vspace{-0.1in}
\paragraph{Forensic sources} Android on-chip and removable flash memory constitute primary forensic sources, both device-wide and app-specific. System-wide sources can provide supplementary information about the underlying Linux kernel activities (via \texttt{dmesg}), system and device-wide app event logging (via \texttt{logcat}), user account audits, running services, device chipset info, cellular and Wi-Fi network activities (via \texttt{dumpsys}) \cite{hoog2011android}. The \texttt{/data/data} sub-tree of the Android file-system inside the \texttt{userdata} partition, along with the \texttt{sdcard} partition, is where it starts to get interesting, with app data typically stored in XML or SQLite files.
Another forensic source typically associated with mobile devices is cloud storage. Given the large multimedia files handled by Android apps, combined with the available storage, cloud storage has become a popular medium for long-term storage, even used seamlessly by apps for regular operation and backups.
App data is increasingly being stored in encrypted form for security and privacy purposes (e.g.\ practically all mainstream messaging apps \cite{Anglano2017telegram}). Beyond the app-level, device-wide disk encryption has seen an evolution across Android versions. Full disk encryption (FDE) has now been replaced by file-based encryption (FBE) in Android 10 \cite{googlefbe}, rendering it more practical and more stable, e.g.\ the alarm clock works even if the screen is locked and a full factory reset is no longer necessary if the device runs out of power before it shuts down properly.

\vspace{-0.1in}
\paragraph{Acquisition and analysis} Evidence acquisition on Android can be carried out using logical or physical imaging \cite{srivastava2015logical}. Simply put, logical acquisition relies on some existing source to parse raw data to decode OS filesystem or app content. Android's filesystem and SQLite parsers are typical examples. App backup utilities and cloud storage interfaces are further examples. On the other hand, physical imaging provides exact bit-for-bit copies of flash memory partitions and can be conducted purely at the hardware level (e.g. through JTAG). All acquisition methods have to deal with Android's security barriers. For software-based acquisition, the barriers range from locked screens to password-protected cloud storage and rooting the device to gain access to \texttt{/data/data/}. Rooting relies upon exploiting some kernel or firmware flashing protocol vulnerability \cite{yang2015new,srivastava2015logical}, or else flashing a custom recovery partition through which to add some root-privileged utility. The latter may get further complicated by locked bootloaders. While hardware-based acquisitions can bypass the above barriers, any form of physical imaging has to deal with FDE and FBE.

The starting point for forensic analysis pretty much depends on what kind of acquisition is performed \cite{hoog2011android}. In this case of physical acquisition, it is necessary to first identify the filesystem concerned, typically EXT and YAFFS, to extract the individual files with possible decryption efforts. This first pass brings the evidence to a state equivalent to a logically acquired one. A typical analysis pass for Android constitutes SQLite file parsing, given its inherent Android support. From this point onwards, decoding of app evidence is pretty much app-specific.
\vspace{-0.1in}
\paragraph{Mobile forensics tools} As such, any mobile forensics tool, e.g. Oxygen Forensics \cite{Oxygen}, or Cellebrite's UFED \cite{Cellebtrite}, can be seen as a collection of acquisition options, equipped with rooting exploits and hardware interfacing cables, passcode brute-forcing methods, along with parsing/analysis modules for filesystems, database, and app data formats. Ancillary analysis features, including timeline generators, can provide a final professional touch to the product.
\subsection{Forensic timelines}\label{sec:forensictimelines}
Forensic timeline generation is widely considered to be the forensic analysis exercise that brings together all the collected evidence. It supports an investigator in reconstructing the hypothesized incident/crime scenario \cite{hargreaves2012automated}. The richer the timelines, the greater the support there is for an investigator to reconstruct an intrusion/crime scene, thereby answering critical questions about an incident.
\begin{figure}
\centering
\includegraphics[page=3,trim = 30mm 5mm 10mm 10mm, clip, width=1\linewidth]{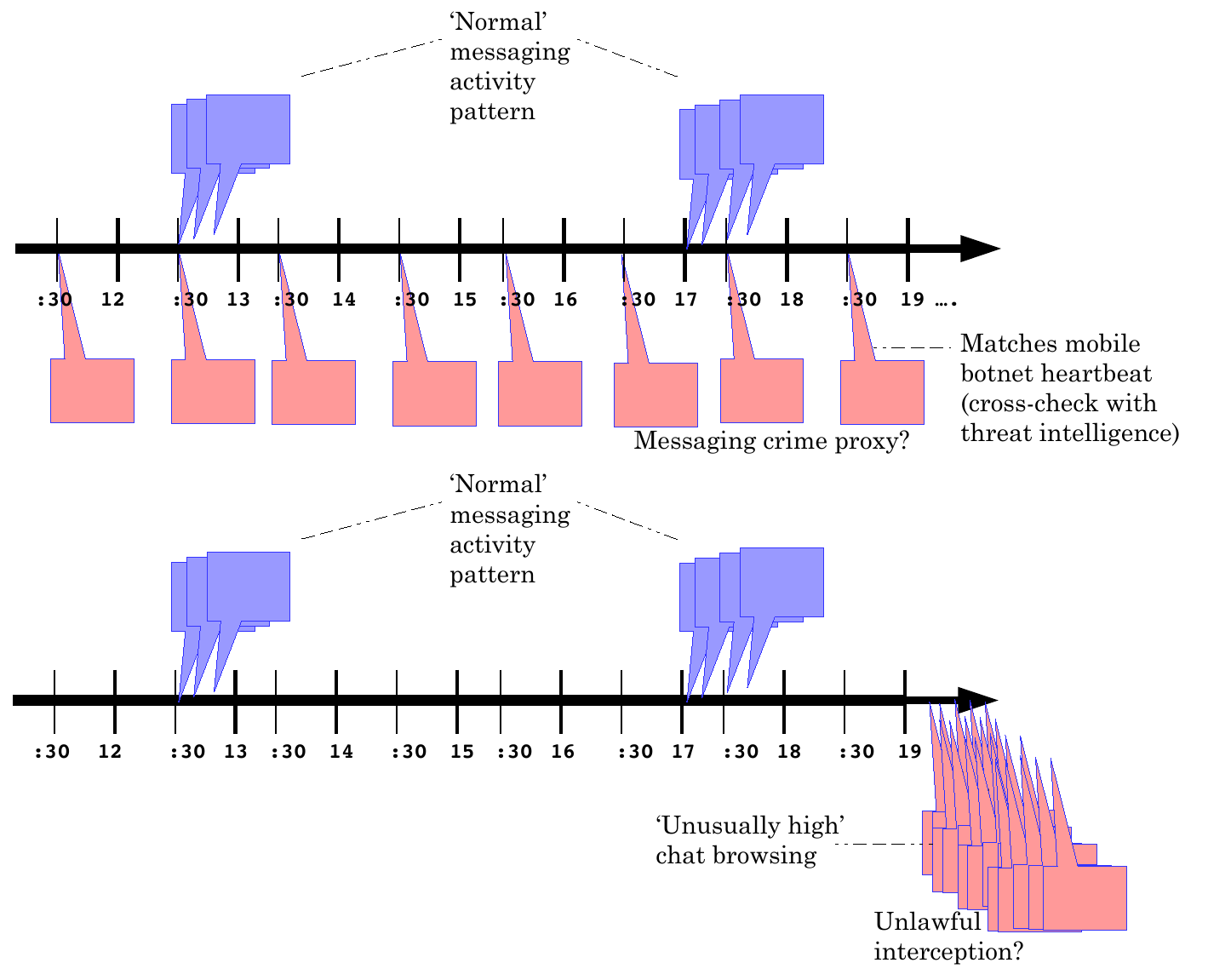}
\caption{Forensic timelines supporting cyber-attack investigations.}
\label{fig:timeline}
\end{figure}

In the case of a messaging hijack, see Figure \ref{fig:timeline} (top), the generated timeline can uncover a pattern matching a cyber threat intelligence feed for mobile botnet activity that could be leveraged for a crime messaging proxy. Else, see Figure \ref{fig:timeline} (bottom), timeline events can be compared to a baseline to identify unusual/suspicious activity. Both approaches can indicate the presence of an ongoing cyberattack, with a comprehensive timeline providing crucial support.

\subsection{Just-in-time Memory Forensics (JIT-MF)}\label{sec:jitmf}
JIT-MF is a framework that enables live process memory forensics in a setting involving attacks that delegate key steps to benign apps, possibly only minimally employing malware. JIT-MF is conceived to be adopted by incident response tools for stock smartphones without breaking any of their security controls. Rather, given that its main purpose is to protect the device owner from cyberattacks and the perpetrators behind them, it assumes the device owner's collaboration. 
Initial experimentation showed JIT-MF's effectiveness to dump ephemeral artefacts at practical overheads. Moreover, while JIT-MF drivers are app- and incident scenario-specific by design, the work involved in defining trigger points only requires a black-box analysis of target apps. An in-depth comprehension only concerns the evidence objects themselves, e.g. the structure of objects representing message objects or messaging activity in SMS or instant messengers. On the downside, pending some form of privacy-preserving computation to be added to JIT-MF, privacy concerns must be addressed solely through procedural rather than technical controls. Limitations also apply in terms of clashes with anti-repackaging measures and dependency on Android RunTime (ART) data structures bound to change between Android versions.

\subsection{Related work}\label{sec:relatedwork}
The topic of Android messaging timelining has already received attention from various works \cite{akinbi2021forensic,Anglano2017telegram,anglano2014forensic}. Their focus, however, is exclusive to content stored on internal flash memory, typically SQLite database schemas and any corresponding encryption and decryption key location. Also, the automated analysis of forensic timelines \cite{CHABOT2015ontol,MOHAMMAD2019mlfortimeline} is orthogonal to our work. JIT-MF's scope is to provide a solution for rendering the acquisition of incident-related evidence as comprehensive as possible, as otherwise no manual or automated process would be able to reconstruct the incident.

JIT-MF relies on a combination of static and dynamic instrumentation of compiled code to implement trigger points as in-line function hooks and the memory dumping process in the form of instrumentation code. Binary instrumentation also underpins various other Android security techniques where it is required to operate within real-time parameters \cite{diamantaris2019reaper,heuser2014asm,chen2017uncovering,li2015real}. Furthermore, JIT-MF considers the temporal aspect of volatile memory forensic collection concerned with the timely collection of ephemeral artifacts. Other temporal aspects that got the attention of similar memory forensics studies concern the memory smearing problem of full device memory dumps \cite{pagani2019introducing} and the extraction of the temporal dimension from static dumps \cite{saltaformaggio2015vcr,ali2019droidscraper}.

In this work, we focus on the attack vector presented by Android accessibility services since it presents an ongoing threat, with multiple recent incidents gaining a world-wide reach \cite{eventbot,defensorid,androidrat}. Crucially, for incident response, the resulting reduced forensic footprint for any attack employing it has also been demonstrated \cite{leguesse2020reducing}. Yet, other attack vectors may also present the same opportunity for LOtL tactics. Zygote and binder infection combined with a rooting exploit \cite{triada}, as well as app-level virtualization frameworks \cite{shi2019jekyll} and third-party library infections \cite{diamantaris2019reaper} provide further attack vectors, resulting in similarly stealthy attacks and for which JIT-MF could be a solution in terms of incident response.

\vspace{-0.15in}
\section{JIT-MF drivers}\label{sec:drivers}
While JIT-MF \cite{bellizzi2021real} defines common steps to be followed by every JIT-MF tool, those aspects that are specific to the investigation scenario/target app pair at hand are described and eventually implemented by JIT-MF drivers. Conceptually, their definition starts with identifying the in-memory evidence objects of interest, which may correspond in some way to attack steps. Next comes trigger point selection, which corresponds to function hooks being placed in native or managed code at the implementation level. Trigger points can be rendered more specific through complementary conditions defined over function arguments. The acquisition method must be defined as either online or offline, specifying whether object carving/parsing is carried out prior to memory dumping or else afterwards. Finally, a sampling strategy is defined to maintain a manageable amount of dumps. Listing \ref{lst:drivertemplate} presents a template generically describing JIT-MF drivers:\\

\begin{lstlisting}[frame=single,numbers=left,caption={JIT-MF driver template},label={lst:drivertemplate},basicstyle=\tiny,mathescape=true, breaklines=true]  
$Driver\_ID$: string
$Scope$: <app, incident_scenario>
/*Attributes*/
$Evidence\_objects$: {<event: string, object_name: string, 
    carve_object_type(), parse_object_type(), {trigger ids}>, ...}
$Acquisition$: {online | offline}
$Triggers$: {<trigger_id: string,<hooked_function_name: string,
    level: {native|rt}, trigger_predicate()>, 
    trigger_callback()>,....}
$Sampling\_strategy$: sampling_predicate()
$Log\_location$: string
$Globals$: {<key,value>, ...}
/*Exposed interface*/
bool $init$(config: {<key,value>, ...}) {
    $set\_globals$(config: {<key,value>, ...});
    for all entries in $Triggers$:
        $place\_native_\_hook$() $\oplus$ $place\_rt\_hook$();
}
/*Internal functions*/
bool $trigger\_predicate_{i}$(params: {<key,value>, ...}) {
    decide on whether to fire the corresponding trigger;
}
void $trigger\_callback_{i}$(thread_context: {<key,value>, ...}) {
    if  $trigger\_predicate_{i}$ && $sampling\_predicate$ return true:
        perform memory forensics on the current app state;
}
[object: address,...] $carve\_object\_type_{j}$(from: address,
    to: address) {
    attempt object carving in the given memory range;
}
[field: type,...] $parse\_object\_type_{j}$(at: address) {
    parse object fields starting at the given address;
}
bool $sampling\_predicate$(thread_context: {<key,value>, ...}) {
    decide on whether to follow up a trigger by a memory dump;
}
\end{lstlisting}
Lines 1-2 identify the driver ($\textit{Driver\_ID}$) and link it with its intended app/incident $\textit{Scope}$. Lines 4-12 enlist a driver's attributes and their types ($attribute: type$), with tuples denoted by $\textit{<>}$, sets by $\textit{\{x,y,z...\}}$, ordered lists by $[]$, key-value pairs by $\textit{<key,value>}$ and enumerations with $\textit{\{val1|val2|...\}}$. Function parameters are identified by the final parenthesis $()$, and these correspond to internal functions in the drivers (lines 20-36). $\textit{Globals}$ is a key-value meant for miscellaneous usage. 

$init()$ presents the only interfaces exposed to the JIT-MF tool environment. It is called during tool initialization and takes care of initializing $\textit{Globals}$ and most importantly sets up the event $\textit{Trigger}s$ by calling $\textit{place\_native|rt\_hook(})$. This function returns a boolean ($\textit{bool}$) indicating success or otherwise. $\textit{Trigger\_predicate()}$ and $\textit{Trigger\_callback()}$ must be defined per entry in $\textit{Triggers}$. Triggers may concern either $native$ or $rt$ function hook, with the latter implying the device's runtime environment, e.g. ART in the case for Android. The same applies for $\textit{carve\_object\_type()}$ and $\textit{parse\_object\_type()}$, which have to be both defined per entry in $\textit{Evidence\_objects}$, at least for online $\textit{Acquisition}$. All these functions require a JIT-MF runtime for their implementation. Listing \ref{lst:driverruntime} presents this runtime assumed by JIT-MF drivers which needs to be catered for by the JIT-MF tool:\\

\begin{lstlisting}[frame=single,numbers=left,caption={JIT-MF driver runtime},label={lst:driverruntime},basicstyle=\tiny,mathescape=true,breaklines=true]  
bool $set/get\_global$(config: <key,value>)
bool $place/remove\_native\_hook$(module!function);
bool $place/remove\_rt\_hook$(namespace.object.method);
[<start: address,end: address, permissions : {---|r--|rw-|rwx|...}, 
    mapped_file: string>,...]  $list\_memory\_segments$();
bool  $set\_memory\_permissions$(segmentbase: address, 
    permissions : {---|r--|rw-|rwx|...});
[byte, ...] $read\_memory$(at: address, length: integer);
bool $dump\_memory\_segment$(from: address, to: address, 
    location: string);
bool $dump\_native\_object$(from: address, to: address, location: string, 
    $carve\_object\_type_{j}$(), $parse\_object\_type_{j}$());
bool $dump\_rt\_object$(namespace.object, $carve\_object\_type_{j}$(), 
    $parse\_object\_type_{j}$);
return_type  $call\_native\_function$(at: address);
return_type  $call\_rt\_function$(namespace.object.method);
bool $append\_log$(path: string,  value: string);
\end{lstlisting}

Lines 1-3 are $\textit{Globals}$ access and $\textit{native$/$rt}$ function-hooking functions called from $\textit{init()}$ and any other driver internal functions as needed. Lines 4-16 are process memory interacting functions, starting off $\textit{list\_memory\_segments()}$ in order to make sure the driver does not attempt to access un-mapped memory, or segments for which it has insufficient permissions. Memory dumping may therefore require adjusting permissions through $\textit{set\_memory\_permissions()}$, as well as checking memory content through $\textit{read\_memory()}$. While for offline $\textit{Acquisition}$, calling $dump\_memory\_segment$ may suffice, for online acquisition the driver is also required to carve objects and parse their fields. $\textit{dump\_native\_object()}$ and $\textit{dump\_rt\_object}$ are utility functions that do just that, taking the appropriate $\textit{carve\_object\_type()}$ and $\textit{parse\_object\_type()}$ callback functions as parameters. Separate $\textit{rt}$ and $\textit{native}$ versions are needed since the $\textit{rt}$ version may leverage calling runtime functions in order to locate the required objects. Similarly, the $\textit{native}$ version may leverage any memory allocators being used to manage native objects. $c\textit{all\_native\_function()}$ and $\textit{call\_rt\_function()}$ functions are utility functions that may be needed by both driver and runtime functions. Finally, $\textit{append\_log()}$ (line 17) is responsible to produce the actual JIT-MF dump to the location specified by the driver's $\textit{Log\_location}$.

\vspace{-0.19in}
\section{Methodology}\label{sec:method}
\vspace{-0.09in}
Before proceeding to demonstrate the value that JIT-MF drivers (section \ref{sec:exp}) add through their role as forensic sources in incident response, we first explain how specific JIT-MF drivers were constructed for the case studies considered. We also describe the methodology used to merge the JIT-MF forensic sources with all other available sources and the eventual resulting timelines of the incidents concerned.

\subsection{JIT-MF driver definition}\label{sec:driver design}
Taking previous experiment results \cite{bellizzi2021real} as guidance for JIT-MF driver definition, white-box analysis of the apps concerned should be restricted to $Evidence\_objects$ identification and implementing their corresponding carving/parsing. Consequently, some form of app code comprehension or reversing is required. On the other hand, $Triggers$ identification should follow a black-box approach. The implication is that the hooked functions can be identified based on calls made to Android APIs, well-known third-party libraries, or even Linux system calls performed by the app, rather than hooking the app's internal functions. The finalized methodology that was adopted in this regard follows:
\begin{itemize}
\item $\textit{Evidence\_objects}$: These objects are identified as those whose presence in memory, in the context of a specific trigger point, implies the execution of some specific app functionality, possibly a delegated attack step. Not all objects are associated with the same level of granularity concerning app events; some objects may be highly indicative of a detailed app event, e.g.\ a message object with an attribute \textit{sent=true|false}, others may only reflect vague app usage across a time period. Therefore when selecting evidence objects, one has to keep in mind how tightly coupled the presence of the objects is with the app functionality that needs to be uncovered. 
\item $\textit{Triggers}$: Taking into account an attack scenario, corresponding target app functionality, and the associated evidence objects, trigger points are selected based on the code that processes the said objects, specifically concerning: \emph{i)} The storing and loading of the objects from storage; \emph{ii)} The transferring of objects over the network; or else \emph{iii)} Any object transformation of some sort (e.g. display on-screen etc.). As yet, we are limited to introducing trigger points only in the \textit{main process} of an app.
\end{itemize}

\subsection{Forensic timeline analysis}\label{sec:ftimeline}

Forensic timeline generation considers all those sources that can shed light on app usage. These range from the device-wide \texttt{logcat} to app-specific sources inside \texttt{/data/data}, as well as inside removable storage which can be found in the \texttt{sdcard} partition and whose mount point is device-specific. Whenever the same data could be obtained from multiple sources, we opt for local device acquisition, rather than cloud or backups,  to facilitate experimentation. These forensic sources, as explained in section \ref{sec:androidforensics}, are representative of those targeted by state-of-the-art mobile forensics tools, typically also requiring device rooting or a combination of hardware-based physical acquisition and content decryption. These baseline sources are complemented by those provided by JIT-MF drivers.

\begin{figure*}
\centering
\includegraphics[page=4,trim = 0mm 0mm 0mm 0mm, clip, width=0.7\linewidth]{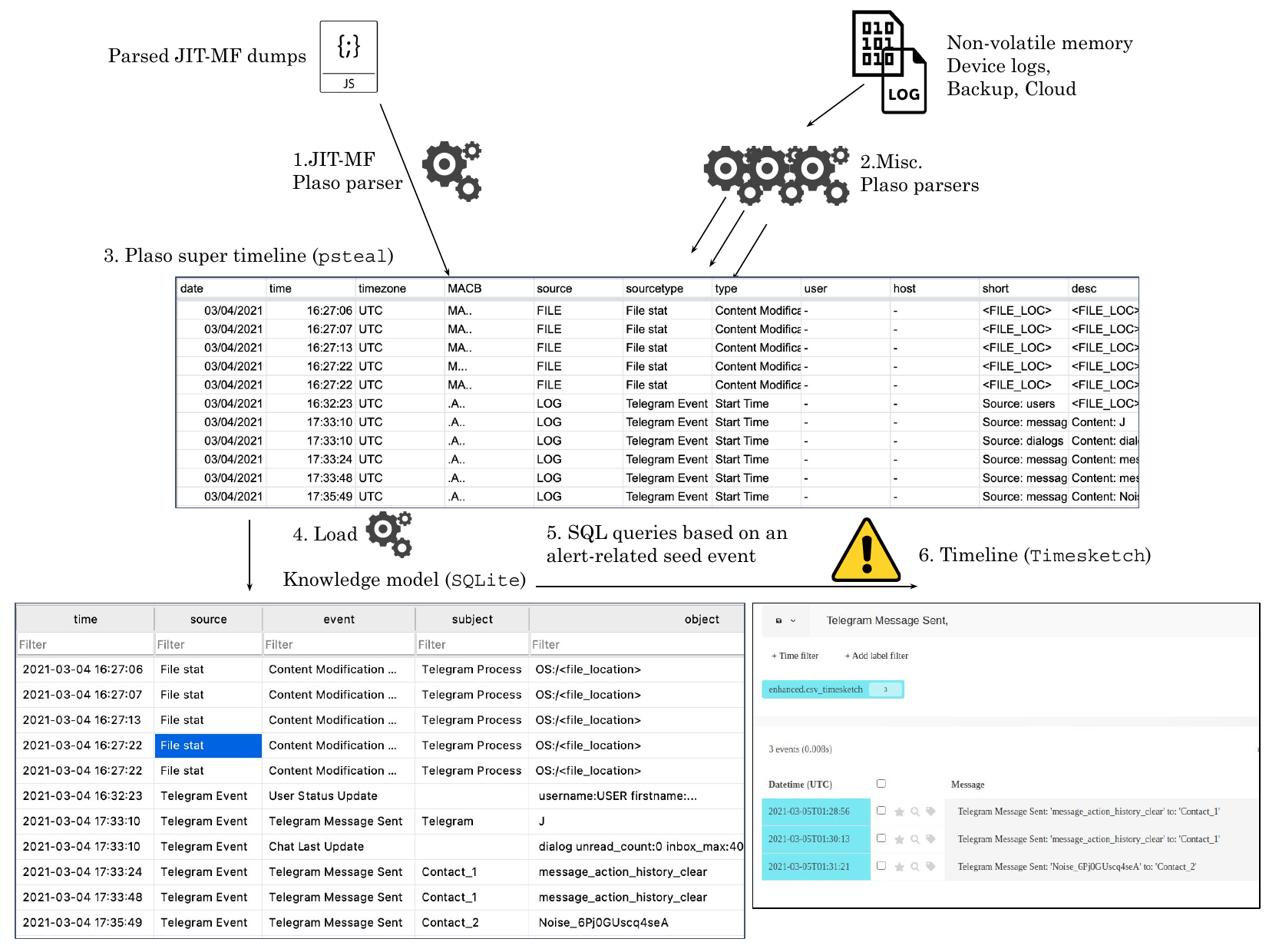}
\caption{The forensic timeline generation processes.}
\label{fig:timelinemethod}
\end{figure*}

Figure \ref{fig:timelinemethod} shows the processes that transform the forensic footprints obtained from the aforementioned forensic sources to finished forensic timelines. This pipeline is based on Chabot et al.'s \cite{chabot2014complete} methodology. It revolves around the creation of a knowledge representation model as derived from multiple forensic sources. It presents a canonical semantic view of the combined sources upon which forensic timeline (or other) analysis can be conducted. This model is populated with scenario events derived from forensic footprints, which are the raw forensic artefacts collected from the different forensic sources. These events are associated with subjects that participate or are affected by the events and the objects acted upon by subjects. Events can subsequently be correlated based on common subjects, objects, as well as temporal relations or expert rule-sets. Relations established by this process correspond either to a relation of composition or causation.

The first three steps in Figure \ref{fig:timelinemethod} consist of forensic footprint extraction. For JIT-MF we refer to a combined dump containing unique, readily carved and parsed memory objects. All sources are decoded and merged as a Plaso \cite{plaso} super timeline using the \texttt{psteal} utility, and for which we developed a JIT-MF Plaso parser. A loader utility was developed for Step 4 that traverses the super timeline and populates the knowledge model, which we store in an \texttt{SQLite} database table. The events in this table correspond to messaging events of some form depending on the forensic source. For example, JIT-MF drivers and messaging backups can pinpoint events at the finest possible level of granularity, indicating whether a specific event is a message send or receive, the recipient/sender. Other sources, such as the file-system source (\texttt{file stat}), can only provide a coarser level of events related to the reading/writing of app-specific messaging database files on the device. We stick to a flat model for this experimentation, with events considered atomic and their associated subjects and objects corresponding to message recipients and content, respectively. Step 5 takes alerts of suspicious activity as input. Alert information based on a seed event is converted into SQL queries that encode the required subject/object/temporal/event type correlations. The query then outputs those events associated with the initial alert. This event sequence provides the timeline for the incident in question, and for which, in step 6, we used \texttt{Timesketch} \cite{timesketch} for visualization.

\section{Experimentation}\label{sec:exp}
To demonstrate the added value that JIT-MF tools provide during incident response, in comparison with current mobile forensics tools, we consider a suite of case studies inspired by real-life accessibility attack scenarios that target messaging apps. Each case study assumes a high-profile target victim ("John"), whose Android mobile device stores confidential data. John makes use of multiple messaging apps which have been forensically enhanced with JIT-MF due to the potentially heightened threats that his device may face. John receives an email to download a free version of an app that he currently pays for on his mobile device. He downloads it and becomes a victim of a long-term stealthy attack.

\noindent\textbf{Setup.} Pushbullet
(v18.4.0), Telegram
(v5.12.0) and Signal
(v5.4.12) are popular SMS and Instant Messaging apps, respectively, used in our case studies. The messaging hijack scenarios considered involve unlawful interception and crime-proxying. For the case studies involving Telegram and Signal, these attacks are carried out using the Android Metasploit attack suite, whereas, for Pushbullet, these attacks are executed through Selenium. Since most state-of-art forensic sources require a rooted device, a rooted emulator is used in our experiment, which also enabled ease of automation. The emulator used was Google Pixel 3XL developer phone running Android 10. The JIT-MF driver runtime was provided by a subset of the Frida\footnote{\url{https://frida.re/docs/android/}} runtime, and JIT-MF drivers were implemented as Javascript code for Frida's Gadget shared library. The JIT-MF drivers used in all case studies have the following attributes: $\textit{Acquisition = online}$, $\textit{Sampling = 1\ in\ 5}$ (active for a second every 5 seconds) and $\textit{Log\_location=/<external\_storage>/jitmflogs}$. Table \ref{tab:forensic_sources} lists the properties of the state-of-the-art forensic sources considered, their method of acquisition and required parsers for populating a Plaso super timeline. Apart from the JIT-MF Plaso parser, additional parsers for the app-specific databases (including write-ahead-log database) and \texttt{logcat} were created.

\noindent\textbf{Case Study Setup.} In each of the case studies: the emulator is started, the target app is instrumented, legitimate traffic (noise) and malicious events are simulated, forensic sources of evidence are collected (both baseline and JIT-MF sources), and timelines are produced based on a knowledge model. The output of these steps are the following generated timelines:
\begin{itemize}
    \item a \textit{Ground Truth Timeline} generated by logging the individual attack steps of the executed accessibility attack.
    \item \textit{Baseline Timelines} generated by querying a knowledge model made up of state-of-the-art forensic sources and
    \item \textit{JIT-MF Timelines} generated by querying a knowledge model made up of both baseline sources and JIT-MF dumps.
\end{itemize}
While there is only one ground truth timeline, multiple JIT-MF and baseline timelines can be created per case study depending on the different seed event correlations. In each case study, the malicious event (attack) is executed three times to ensure that the ground truth timeline contains more than one event. Noise is generated with respect to this value. Some of the attacks used in these case studies make use of rate-limited API calls to a server backend, which only allows 150 consecutive calls to be made from the same device. Since an attack is executed three times per case study, and the API call limit is 150, each case study is simulated 50 times --- each time obtaining the timelines above.


\noindent\textbf{Timeline Comparison.} The \textit{JIT-MF Timeline} and \textit{Baseline Timeline} are compared to the \textit{Ground Truth Timeline} based on: \emph{i)} completeness of timeline, i.e. lack of missing events; \emph{ii)} accuracy of the timelines with respect to the sequence in which the events happened and the difference between the recorded time of an event in the ground truth timeline and the JIT-MF timeline.
Preliminary runs showed that baseline forensic sources could provide different metadata depending on the app in use. Therefore, the matching criteria for a matched event between either of the generated timelines and the ground truth timeline are adjusted in the case studies to benefit from the evidence typically found in baseline forensic sources.

\renewcommand{\arraystretch}{0.7}
\begin{table*}[htpb!]
\centering
\scriptsize
\caption{Forensic Sources and Parsers.}
\begin{tabular}{|>{\centering\arraybackslash}m{0.6cm}|>{\centering\arraybackslash}m{4.2cm}|>{\centering\arraybackslash}m{0.8cm}|>{\centering\arraybackslash}m{3.2cm}|>{\centering\arraybackslash}m{1.4cm}|>{\centering\arraybackslash}m{0.9cm}|>{\centering\arraybackslash}m{1.6cm}|}
\hline
\centering \textbf{Case Study} &  \textbf{Location on device} & \textbf{Source type} & \textbf{Contents} & \textbf{Acquisition \& Decoding} & \textbf{Requires rooting} & \textbf{Plaso parsers} \\ \hline
A,D & \url{/data/org.telegram.messenger/.../cache4.db} & Baseline & Telegram database & \texttt{adb pull}, Teleparser\footnote{\url{https://github.com/RealityNet/teleparser}} & Yes & Teleparser parser \\ \hline
A,D & \url{/data/org.telegram.messenger/.../cache4.db-wal} & Baseline & Latest changes to Telegram's database & \texttt{adb pull}, Walitean\footnote{\url{https://github.com/xperylabhub/walitean.git}} & Yes & Walitean parser \\ \hline
B,E & \url{/<removable_storage>/.../signal.backup} & Baseline & Signal backup database & Signal DB decryptor\footnote{\url{https://github.com/xeals/signal-back}} & Yes & Signal database parser \\ \hline
D,F & \url{/data/data/com.android.providers.telephony/.../mmssms.db} & Baseline & SMS database & \texttt{adb pull} & Yes & \texttt{mmssms.db} Plaso parser \\ \hline
D,F & \url{/data/data/com.pushbullet.android/.../pushes.db} & Baseline & Pushbullet message database & \texttt{adb pull} & Yes & Pushbullet parser \\ \hline
A-F & \url{/data/<app_pkg_name>/*} & Baseline & App specific files, cache files & \texttt{adb pull} & Yes & File stat Plaso parser \\ \hline
A-F & \url{/<removable_storage>/<app_pkg_name>} & Baseline & Media files & \texttt{adb pull} & No & File stat Plaso parser \\ \hline
A-F & \texttt{logcat} & Baseline & System logs & \texttt{adb logcat} & No & Logcat parser \\ \hline
A-F & \url{/<removable_storage>/jitmflogs} & JIT-MF & JIT-MF memory dumps & \texttt{adb pull} & No & JIT-MF parser \\ \hline
\end{tabular}%
\label{tab:forensic_sources}
\end{table*}

\subsubsection{Telegram Crime-Proxy}\label{sec:casestudies}

\noindent\textbf{Accessibility attack.} An accessibility attack targets John's Telegram app and is used by an attacker to send messages to a co-conspirator going by the username "Alice" on Telegram. The attacker misuses the victim's Telegram app to send messages to "Alice" and instantly deletes them.

\noindent\textbf{Setup.} John makes use of his Telegram app regularly to communicate with his family and friends. He sends six Telegram messages to his relatives before entering a meeting, then goes silent. The attacker notices the decrease in Telegram activity and decides to use this time to communicate with "Alice" three times. He waits 10 - 20 seconds (randomly generated using \texttt{rand}) every time before messaging "Alice". The attacker tries to execute the attack as quickly as possible to retain stealth but gives an allowance of 10 seconds within the attack to allow for any delays within the app. John continues using Telegram thereafter and sends six messages to his friend. John's messages take this form: $Noise\_<Random10-100-letters>$ whereas those sent by the attacker are similar to $Sending\_Attack\_\#Iteration$. 

\noindent\textbf{Investigation.}
John notices a new chat on his phone with the username "Alice" with no messages. He brushes it off but is contacted later that week by investigators who told him that his phone was used to send messages containing the specific keywords. He takes his phone to be examined. His phone is already equipped with a JIT-MF driver that has the below attributes:\\
\noindent\textbf{\textit{Evidence\_objects}}: \{<"Telegram Message Sent",\\ "\texttt{org.telegram.messenger.MessageObject}", \{"TG\_CP\}">\}\\
\noindent\textbf{\textit{Triggers}}: \{<"TG\_CP<", "\texttt{send}", native, network>\!>\}\\
This attack step involves the sending of a message over the network. Therefore the selected trigger point is the \texttt{send} system call, and the evidence object is the Telegram message itself.

The seed event is generated based on the alert flagged, which gives the investigators three possible starting points to use when formulating the queries to be executed on the different knowledge models.\\
\noindent\textbf{\textit{Seed Event}}: \emph{Subject: Alice, Object: *specific keywords*, Event type: Message Sent, Time: last 7 days}\\
\noindent\textbf{\textit{Matching criteria}}: The criteria for an event in the baseline or enhanced timelines to match the ground truth timeline is the presence of the specific message content that was sent within the event.

\subsubsection{Signal Crime-Proxy}

\noindent\textbf{Accessibility attack.} An accessibility attack targets John's Signal app and is used by an attacker to send messages to a co-conspirator going by the username "Alice" on Signal. The attacker misuses the victim's Signal app to send messages to "Alice" and instantly deletes them. 

\noindent\textbf{Setup.} This case study is identical to the one described in the previous section.

\noindent\textbf{Investigation.}
John's phone is already equipped with a JIT-MF driver that has the below attributes:\\
\noindent\textbf{\textit{Evidence\_objects}}: \{<"Signal Message Sent","\texttt{org.\\thoughtcrime.securesms.conversation.\\ConversationMessage}", \{"SIG\_CP"\}>\}\\
\noindent\textbf{\textit{Triggers}}:\{<"SIGNAL\_CP",<"\texttt{write}",native,storage>\!>\}\\
Similar to the previous case study, the evidence object is the Signal message itself. Signal does not make use of the \texttt{send} system call however when sending a message. The \texttt{write} system call is used as a trigger point, which writes to the local database and over the network.

\noindent\textbf{\textit{Seed Event}}: \emph{Subject: Alice, Object: *specific keywords*, Event type: Message Sent, Time: last 7 days}\\
\noindent\textbf{\textit{Matching criteria}}: An event stating that a message was sent to Alice's number.

\subsubsection{Pushbullet Crime-Proxy.}\label{sec:pushbullet_cp}
\paragraph{Accessibility attack} John's Facebook credentials are stolen by an attacker using a phishing accessibility attack akin to Eventbot\cite{eventbot}. The attacker uses the stolen credentials to proxy SMSs, through John's Pushbullet app, from his web browser.

\noindent\textbf{Setup.} John does not use SMS functionality on his phone but is aware that he receives many advertisement messages. John receives six ad messages prior to entering a meeting. The attacker notices the decrease in activity and decides to use this time to communicate with "Alice" three times. He waits 10 - 20 seconds (randomly generated using \texttt{rand}), then opens his browser and sends three messages to "Alice". Messages received by John take this form: $Noise\_<Random10-100-letters>$ whereas those sent by the attacker are similar to this: $Sending\_Attack\_\#Iteration$. 

\noindent\textbf{Investigation.}
John receives a hefty bill at the end of the month from his telephony provider, attributing most of the cost to message sending. He notices a new number that is not on his contact list and takes his phone to be examined. His phone is already equipped with a JIT-MF driver that has the below attributes:\\
\noindent\textbf{\textit{Evidence\_objects}}: \{<"Message Sent","\texttt{org.json.\\JSONObject}",\{"PB\_CP"\}>\}\\
\noindent\textbf{\textit{Triggers}}: \{<"PB\_CP",<"\texttt{write}",native,storage>\!>\}\\
Pushbullet stores message objects in JSON structures. A \texttt{write} system call trigger point occurs when a message is sent, at which point the process memory contains the message sent, stored in JSON.

\noindent\textbf{\textit{Seed Event}}: \emph{<Subject: *suspicious number*, Event type: Message Sent>, Time: last month}\\
\noindent\textbf{\textit{Matching criteria}}: A message sent to the suspicious number.

\subsubsection{Telegram Unlawful Interception}

\noindent\textbf{Accessibility attack.} An accessibility attack targets John's Telegram app and is used by an attacker to intercept messages sent to the username "CEO" (John's boss - with whom confidential data is shared). The attacker misuses John's Telegram app to grab messages exchanged with "CEO" and Telegram.

\noindent\textbf{Setup.} John makes use of his Telegram app regularly to communicate with his CEO. John sends messages to his CEO multiple times during the day but goes silent during three meetings. The attacker notices the decrease in Telegram activity and decides to use this time to spy on John's correspondence with his CEO. He waits 10 - 20 seconds (randomly generated using \texttt{rand}), then opens Telegram, loads the "CEO" chat, intercepts the messages loaded on the screen then closes the app quickly. Messages sent by John take this form: $Confidential\_<Random10-100-letters>$.

\noindent\textbf{Investigation.}
John's phone is already equipped with a JIT-MF driver that has the below attributes:\\
\noindent\textbf{\textit{Evidence\_objects}}: \{<"Telegram Chat Intercepted",\\"\texttt{org.telegram.messenger.MessageObject}",\{"TG\_SP"\}>\}\\
\noindent\textbf{\textit{Triggers}}: \{<"TG\_SP",<"\texttt{recv}",native,network>\!>\}\\
In the case of an unlawful interception attack, one of the attack steps involves the reading of a message, therefore the evidence object is the message itself. Since Telegram is a cloud-based app, some messages are stored on the device, and others are loaded and received from cloud storage over the network. Therefore the selected trigger point is the \texttt{recv} system call. 

\noindent\textbf{\textit{Seed Event}}: \emph{Subject: CEO, Object: *confidential message*, Event type: Message Read/Loaded/Chat activity, Time: date of message sent/received}\\
\noindent\textbf{\textit{Matching criteria}}: An event type indicating chat activity, loading or reading of "CEO" messages. The message object itself does not correspond directly to an attack step. That is, the message object in memory does not contain metadata about whether it was read but rather that it was either sent or received at some point. JIT-MF forensic sources identify a \textit{chat interception event} instead as multiple message objects exchanged with the same contact, all having been dumped at the same timestamp. Furthermore, the timestamp of these events must occur in the database any time after the sending time to avoid including data related to when the message was initially sent or received.

\subsubsection{Signal Unlawful Interception}

\noindent\textbf{Accessibility attack.} An accessibility attack targets John's Signal app and is used by an attacker to intercept messages sent to the username "CEO". The attacker misuses John's Signal app to open a confidential chat with the username "CEO" and grabs the messages that appear on the screen. Finally, the attacker closes Signal.

\noindent\textbf{Setup.} This case study is identical to the previous one. 

\noindent\textbf{Investigation.}
John's phone is already equipped with a JIT-MF driver that has the below attributes:\\
\noindent\textbf{\textit{Evidence\_objects}}: \{<"Signal Chat Intercepted", "\texttt{org.thoughtcrime.securesms.conversation.\\ConversationMessage}",\{"SIGNAL\_SP"\}>\}\\
\noindent\textbf{\textit{Triggers}}: \{<"SIG\_SP",<"\texttt{open}",native,storage>\!>\}\\
Similar to the previous case study, the evidence object is the Signal message that was intercepted. Signal is not a cloud based app and uses solely on-device storage. Therefore we select the \texttt{open} system call which is used to open the database file from which messages are loaded to be read.

\noindent\textbf{\textit{Seed Event}}: \emph{Subject: CEO, Object: *confidential message*, Event type: Message Read/Loaded/Chat activity, Time: date of message sent/received}\\
\noindent\textbf{\textit{Matching criteria}}: As previous case study.\\

\renewcommand{\arraystretch}{0.7}
\begin{table*}[htbp!]
\scriptsize
\centering
\caption{Timeline comparison.}
\begin{tabularx}{\linewidth}{!>{\centering\arraybackslash}P{1.4cm}!>{\centering\arraybackslash}P{1.885cm}!>{\centering\arraybackslash}P{1.2cm}|>{\centering\arraybackslash}P{1.2cm}|>{\centering\arraybackslash}P{2.9cm}!>{\centering\arraybackslash}P{1.2cm}|>{\centering\arraybackslash}P{1.2cm}|>{\centering\arraybackslash}P{2.9cm}!}
\thickhline
\multirow{4}{=}{\centering\textbf{Case Study}}  & \multirow{4}{=}{\centering\textbf{Seed event - Correlation}} & \multicolumn{3}{c!}{\textbf{Baseline}} & \multicolumn{3}{c!}{\textbf{JIT-MF Timeline}} \\ \cline{3-8}
& & \multirow{3}{=}{\centering\textbf{Recall}} & \multirow{3}{=}{\centering\textbf{Precision}} & \multirow{3}{=}{\centering\textbf{Timeline difference (Jaccard dissimilarity)}} & \multirow{3}{=}{\centering\textbf{Recall}} & \multirow{3}{=}{\centering\textbf{Precision}} & \multirow{3}{=}{\centering\textbf{Timeline difference (Jaccard dissimilarity)}} \\ 
& & & & & & & \\
& & & & & & & \\ \thickhline
\multirow{3}{=}{\centering A} & Subject & 0 & - & 1 & 0.98 & 1 & 0.02 \\ \cline{2-8}
& Object & 1 & 0.66 & 0 & 1 & 0.66 & 0 \\ \cline{2-8}
& Event Type & 1 & 0.01 & 0 & 1 & 0.01 & 0 \\ \thickhline
\multirow{3}{=}{\centering B} & Subject & 1 & 0.07 & 0 & 1 & 0.06 & 0 \\ \cline{2-8}
& Object & 0 & - & 1 & 0.87 & 1 & 0.13 \\ \cline{2-8}
& Event Type & 1 & 0.11 & 0 & 1 & 0.07 & 0 \\ \thickhline
\multirow{2}{=}{\centering C} & Subject & 1 & 1 & 0 & 1 & 1 & 0 \\ \cline{2-8}
& Event Type & 1 & 0.23 & 0 & 1 & 0.23 & 0 \\ \thickhline
\multirow{3}{=}{\centering D} & Subject & 0 & - & 1 & 0.49 & 0.46 & 0.51 \\ \cline{2-8}
& Object & 0 & - & 1 & 0.49 & 0.45 & 0.51 \\ \cline{2-8}
& Event Type & 0 & - & 1 & 0.49 & 0.45 & 0.51 \\ \thickhline
\multirow{3}{=}{\centering E} & Subject & 0.99 & 0.97 & 0.01 & 0.99 & 0.21 & 0.01 \\ \cline{2-8}
& Object & 0 & - & 1 & 0.58 & 0.23 & 0.42 \\ \cline{2-8}
& Event Type & 0.13 & 0.01 & 0.87 & 0.63 & 0.02 & 0.37 \\ \thickhline
\multirow{3}{=}{\centering F} & Subject & 0 & 0 & 1 & 0 & 0 & 1 \\ \cline{2-8}
& Object & 0 & 0 & 1 & 0 & 0 & 1 \\ \cline{2-8}
& Event Type & 0 & - & 1 & 0.02 & 1 & 0.98 \\ \thickhline
\end{tabularx}
\label{tab:results_table}
\end{table*}

\subsubsection{Pushbullet Unlawful Interception.}\label{sec:pushbullet_sp}
\paragraph{Accessibility attack} John's Facebook credentials are stolen by an attacker using a phishing accessibility attack. The attacker now has access to any messages sent or received by John through a syncing event on John's phone.
\vspace{-0.22in}
\paragraph{Setup.} John makes use of his SMS app regularly to communicate with his CEO. John sends messages to his CEO multiple times during the day but goes silent during three meetings. Unbeknownst to him, the attacker is immediately intercepting all of John's ongoing SMS activity.

\noindent\textbf{Investigation.}
John's phone is already equipped with a JIT-MF driver that has the below attributes:\\
\noindent\textbf{\textit{Evidence\_objects}}: \{<"Chat Intercepted", "\texttt{org.\\json.JSONObject}",\{"PB\_SP"\}\}\\
\noindent\textbf{\textit{Triggers}}: \{<"PB\_SP",<"\texttt{android.content.\\Intent.createFromParcel}",rt,network>\!>\}\\
Unlike Telegram and Signal, Pushbullet spawns several sub-processes to sync activity generated on the device with that stored in the cloud. While in Case Study C\ref{sec:pushbullet_cp} the attack involves a level of interaction with the device (since the SMS has to be sent from the device after receiving an instruction from the browser), in this case, any message sent or received is assumed to automatically have been intercepted. The trigger point selected is one of the Android API calls used by the Pushbullet to sync sent/received messages via Firebase.
The only evidence object, related to an attack step, that can be retrieved from memory for this case study, is a JSON object containing "push" event metadata which indicates message content has been synced.\\
\noindent\textbf{\textit{Seed Event}}: \emph{Subject: CEO, Object: *confidential message*, Event type: Message Read/Loaded/Chat activity, Time: date of message sent/received}\\
\noindent\textbf{\textit{Matching criteria}}: As previous case study.\\

\subsection{Results}
Table \ref{tab:results_table} shows a comparison between the generated JIT-MF timelines and Baseline timelines, per seed event correlation, to the ground truth timeline. The generated timelines included events unrelated to the attack steps (noise); therefore, \textit{precision and recall} were used.
Precision is a value between 0 and 1, which denotes the average relevant captured events. The higher the value, the larger the portion that attack steps make up of the timeline, i.e. little noise was present. Where the value is '\textit{-}', no events were captured. Recall denotes how many of the executed attack steps were uncovered. Similarly, the higher the value between 0 and 1, the more attack steps that were captured. Timeline difference was calculated using \textit{Jaccard dissimilarity} on the set of true events uncovered by the generated timelines and the ground truth timeline. In this case, the higher the value between 0 and 1, the more dissimilar the generated timeline is to the ground truth.

\noindent\textbf{Primary contributors to timeline similarity.} The timeline difference values in the table show that overall JIT-MF timelines are \emph{at least} as similar to the ground truth as baseline timelines. While the dissimilarity for the baseline timelines varies substantially \textit{within a single case study}, this is not the case for JIT-MF timelines whose distance from ground truth remains roughly the same. Since JIT-MF forensic sources include finer-grained evidence (message content, recipient, date…), the chosen seed event correlation has little to no effect on the output timeline. In contrast, evidence from baseline sources is not as rich, with correlation becoming a critical factor affecting the resulting timelines. Due to the finer-grained metadata available in JIT-MF forensic sources, we can say that even in the scenarios where JIT-MF timelines are equivalent to the baseline in event sequences, these can provide the investigator with richer timelines through more informative events.

The table also shows that JIT-MF timelines are more similar to the ground truth in the case of unlawful interception (case studies D-F) in comparison with the baseline sources, which do not include enough evidence pointing to message reading or browsing chat activity. 

\noindent\textbf{Primary contributors to timeline dissimilarity.} JIT-MF timelines were most dissimilar from the ground truth in the last case study. The main differences from the other case studies here were: i) Many of the app's functionality was delegated to sub-processes that were not instrumented,  and ii) The evidence object defined in the JIT-MF driver was a coarser-grained object (JSON object containing "push" event). Both of these limitations in JIT-MF's driver implementation contributed to a JIT-MF timeline whose gain on the baseline timeline was minimal, with regard to ground truth timeline similarity.

Furthermore, while JIT-MF timelines are more similar to the ground truth timeline than baseline timelines in cases involving unlawful interception (D-F), they are less similar to the ground truth timelines when compared to JIT-MF timelines obtained for the crime proxy case studies (A-C). The difference between these sets of case studies is that in crime proxy scenarios, the evidence object defined in the JIT-MF driver is the fine-grained message object that contains metadata tightly linked to the event itself. In unlawful interception scenarios, we are after coarser-grained events (an indication of a chat being intercepted/synced) since key objects in memory are either not present or do not contain indicative metadata of the ongoing event.

\noindent\textbf{JIT-MF timeline sequence accuracy.} Using \textit{Kendall Tau distance}, we were able to conclude that the sequence of captured events in JIT-MF timelines (containing only ground truth events) is always identical to that in the ground truth timeline. Additionally, the standard deviation between the time of the events logged in the ground truth timelines and that logged in JIT-MF timelines deviates on average by at most 62s.

\noindent\textbf{Performance overheads.} The practical overheads recorded in the initial JIT-MF study \cite{bellizzi2021real} were confirmed. With JIT-MF drivers enabled, only an average increase of 0.5s was registered in Pushbullet turnaround times for SMS operations, as observed from the web browser's Javascript console. Janky frames\footnote{\url{https://developer.android.com/topic/performance/vitals/render}} is an indicator of non-smooth user interactions with GUI apps. With JIT-MF drivers enabled Telegram and Signal had an average increase of 1.59\% and 3.53\% of Janky frames, respectively; that is, the performance penalty overall was less than 4\%.

\section{Discussion}\label{sec:discussion}
\noindent\textbf{Provision of context for evidence objects by trigger points.}  The selection of trigger points is not only crucial to solving the problem of timely dumping ephemeral evidence objects in memory. Trigger points also provide the necessary context for a dumped evidence object. Results show that JIT-MF tools are most successful in generating a quasi-identical timeline to the ground truth timeline when both trigger point \emph{and} the evidence object are tightly linked to the attack step that needs to be uncovered. For instance, a \texttt{send} system call and a message object are directly linked to an attack step involving an attacker misusing a victim’s phone to proxy a message.

\noindent\textbf{Level of event granularity associated with in-memory objects.}
The level of granularity for the events associated with in-memory objects is not always enough for accurate association with specific app use/misuse events.
In the case of coarser-grained events, due to the lack of metadata within the collected evidence, we noticed that, while still present, the additional information as compared to baseline timelines diminishes considerably. This seems to be the root cause for a diminishing return when we compared crime-proxy timelines with those obtained for unlawful interception.


\noindent\textbf{Further evolution of JIT-MF.}
JIT-MF operates under the assumption that by targeting evidence objects representing core app functionality (e.g.\ message objects in messaging apps), we may uncover relevant attack events. However, the issues mentioned above emerge as a result of this. An approach that merits further investigation is to produce a JIT-MF driver such that: i) Selected trigger points could provide more context to the evidence object being dumped and ii) Evidence objects can be more tightly coupled with app use/misuse. 
This would mean that although a black-box approach to selecting trigger points is successful in dumping ephemeral objects in memory, a white-box approach might be preferred, particularly in scenarios where the attack step does not have fine-grained metadata related to ongoing activity, necessary to deduce ongoing actions. Therefore the evidence object itself requires further context. Deeper app analysis may also be required when selecting an evidence object more suited to target specific attack steps, which becomes even more challenging when addressing obfuscated code where the implementation of an object within the app is obscure.   

\noindent\textbf{Threats to validity.} 
Except for Pushbullet, the two forensic sources making up the baseline in four of the six case studies are \texttt{logcat} and write-ahead log files containing incomplete forensic footprints and are subject to frequent log rotation. When performing longer runs with more iterations of the attack component, we noticed that out of 250 deleted messages, the write-ahead log file could only expose 20 unique messages. The same argument can be made for \texttt{logcat}. Since these logs are not captured in a timely and permanent manner, they would be futile in the case of a long-term stealthy attack. The same point cannot be said for Pushbullet based on the obtained results. However, a slightly more sophisticated malware consisting of a second stage which deletes sent messages on the device would expose the limitations of the existing baseline. On the other hand, a JIT-MF-based tool can timely capture and store the evidence object for the possible forensic analysis required at a later stage. A sampling property enables the tool to reduce the amount of storage occupied. However, if the device is running low on storage, then a JIT-MF based tool would run into the same issues of evidence availability as the baseline sources.

Finally, the app and incident scenarios considered in this experiment are only related to messaging hijacks. Other apps may still be subject to issues that require further experimentation.
\section{Conclusions}\label{sec:concl}
Stealthy Android malware based on accessibility has made it difficult for digital forensic investigators to uncover the attack steps involved in an incident. Previous work has shown that using a JIT-MF tool is crucial in dumping key ephemeral in-memory objects that can be linked to attack steps. In this paper, we demonstrate the extent to which JIT-MF tools can enrich forensic sources by comparing the enhanced timelines generated by JIT-MF tools to timelines generated by existing baseline sources. Results show that while JIT-MF does improve on timelines generated by baseline sources without compromising the security of the device, the information gain obtained in the timeline depends on the level of granularity of the attack step linked to the identified evidence object.

\section{Acknowledgements}
This work is supported by the LOCARD Project under Grant H2020-SU-SEC-2018-832735.

\Urlmuskip=0mu plus 1mu
\bibliographystyle{apalike}
{\small
\bibliography{jitmf_timeline}}

\begin{thebibliography}{}

\bibitem[Cel, ]{Cellebtrite}
Cellebtrite ufed.
\newblock \url{https://www.cellebrite.com} Accessed: 24.03.2021.

\bibitem[Oxy, ]{Oxygen}
Oxygen forensics.
\newblock \url{https://www.oxygen-forensic.com/en/} Accessed: 24.03.2021.

\bibitem[Akinbi and Ojie, 2021]{akinbi2021forensic}
Akinbi, A. and Ojie, E. (2021).
\newblock Forensic analysis of open-source {XMPP}/{J}abber multi-client instant
  messaging apps on {A}ndroid smartphones.
\newblock {\em SN Applied Sciences}, 3(4):1--14.

\bibitem[Ali-Gombe et~al., 2019]{ali2019droidscraper}
Ali-Gombe, A., Sudhakaran, S., Case, A., and Richard~III, G.~G. (2019).
\newblock Droid{S}craper: a tool for {A}ndroid in-memory object recovery and
  reconstruction.
\newblock In {\em $RAID$}, pages 547--559.

\bibitem[Ali-Gombe et~al., 2020]{ali2020app}
Ali-Gombe, A., Tambaoan, A., Gurfolino, A., and Richard~III, G.~G. (2020).
\newblock App-agnostic post-execution semantic analysis of {A}ndroid in-memory
  forensics artifacts.
\newblock In {\em ACSAC}, pages 28--41.

\bibitem[Anglano, 2014]{anglano2014forensic}
Anglano, C. (2014).
\newblock Forensic analysis of {W}hats{A}pp messenger on {A}ndroid smartphones.
\newblock {\em Digital Investigation}, 11(3):201--213.

\bibitem[Anglano et~al., 2017]{Anglano2017telegram}
Anglano, C., Canonico, M., and Guazzone, M. (2017).
\newblock Forensic analysis of telegram messenger on android smartphones.
\newblock {\em Digital Investigation}, 23:31--49.

\bibitem[Bellizzi et~al., 2021]{bellizzi2021real}
Bellizzi, J., Vella, M., Colombo, C., and Hernandez-Castro, J. (2021).
\newblock Real-time triggering of {A}ndroid memory dumps for stealthy attack
  investigation.
\newblock In {\em NordSec}, pages 20--36.

\bibitem[Bhatia et~al., 2018]{bhatia2018tipped}
Bhatia, R., Saltaformaggio, B., Yang, S.~J., Ali-Gombe, A.~I., Zhang, X., Xu,
  D., and Richard~III, G.~G. (2018).
\newblock Tipped off by your memory allocator: {D}evice-wide user activity
  sequencing from {A}ndroid memory images.
\newblock In {\em NDSS}.

\bibitem[{Campbell, Christopher and Graeber, Matthew}, 2013]{lotl}
{Campbell, Christopher and Graeber, Matthew} (2013).
\newblock {L}iving {O}ff the {L}and: {A} {M}inimalist’s {G}uide to {W}indows
  {P}ost-{E}xploitation.
\newblock \url{http://www.irongeek.com}.
\newblock Accessed: 24.03.2021.

\bibitem[Case et~al., 2020]{case2020hooktracer}
Case, A., Maggio, R.~D., Firoz-Ul-Amin, M., Jalalzai, M.~M., Ali-Gombe, A.,
  Sun, M., and Richard~III, G.~G. (2020).
\newblock Hooktracer: {A}utomatic detection and analysis of keystroke loggers
  using memory forensics.
\newblock {\em Computers \& Security}, 96:101872.

\bibitem[Case and Richard~III, 2017]{case2017memory}
Case, A. and Richard~III, G.~G. (2017).
\newblock Memory forensics: The path forward.
\newblock {\em Digital Investigation}, 20:23--33.

\bibitem[Chabot et~al., 2014]{chabot2014complete}
Chabot, Y., Bertaux, A., Nicolle, C., and Kechadi, M.-T. (2014).
\newblock A complete formalized knowledge representation model for advanced
  digital forensics timeline analysis.
\newblock {\em Digital Investigation}, 11:S95--S105.

\bibitem[Chabot et~al., 2015]{CHABOT2015ontol}
Chabot, Y., Bertaux, A., Nicolle, C., and Kechadi, T. (2015).
\newblock An ontology-based approach for the reconstruction and analysis of
  digital incidents timelines.
\newblock {\em Digital Investigation}, 15:83--100.

\bibitem[Chen et~al., 2017]{chen2017uncovering}
Chen, J., Wang, C., Zhao, Z., Chen, K., Du, R., and Ahn, G.-J. (2017).
\newblock Uncovering the face of android ransomware: Characterization and
  real-time detection.
\newblock {\em IEEE TIFS}, 13(5):1286--1300.

\bibitem[Cosic and Baca, 2010]{cosic2010framework}
Cosic, J. and Baca, M. (2010).
\newblock A framework to (im) prove ``chain of custody'' in digital
  investigation process.
\newblock In {\em CECIIS}, page 435.

\bibitem[Diamantaris et~al., 2019]{diamantaris2019reaper}
Diamantaris, M., Papadopoulos, E.~P., Markatos, E.~P., Ioannidis, S., and
  Polakis, J. (2019).
\newblock Reaper: real-time app analysis for augmenting the {A}ndroid
  permission system.
\newblock In {\em ACM CODASPY}, pages 37--48.

\bibitem[Google, a]{googlefbe}
Google.
\newblock File-based encryption.
\newblock \url{https://source.android.com/security/encryption/file-based}
  Accessed: 24.03.2021.

\bibitem[Google, b]{timesketch}
Google.
\newblock Timesketch: forensic timeline analysis.
\newblock \url{https://github.com/google/timesketch} Accessed: 24.03.2021.

\bibitem[Gu{\dh}j{\'o}nsson, 2010]{plaso}
Gu{\dh}j{\'o}nsson, K. (2010).
\newblock Mastering the super timeline with log2timeline.
\newblock {\em SANS Institute}.

\bibitem[Hargreaves and Patterson, 2012]{hargreaves2012automated}
Hargreaves, C. and Patterson, J. (2012).
\newblock An automated timeline reconstruction approach for digital forensic
  investigations.
\newblock {\em Digital Investigation}, 9:S69--S79.

\bibitem[Heuser et~al., 2014]{heuser2014asm}
Heuser, S., Nadkarni, A., Enck, W., and Sadeghi, A.-R. (2014).
\newblock {ASM}: A programmable interface for extending {A}ndroid security.
\newblock In {\em USENIX}, pages 1005--1019.

\bibitem[Hoog, 2011]{hoog2011android}
Hoog, A. (2011).
\newblock {\em Android forensics: investigation, analysis and mobile security
  for Google Android}.

\bibitem[{Kaspersky}, 2016]{triada}
{Kaspersky} (2016).
\newblock {T}riada: organized crime on {A}ndroid.
\newblock \url{https://www.kaspersky.com/blog/triada-trojan/11481}.
\newblock Accessed: 24.03.2021.

\bibitem[Leguesse et~al., 2020]{leguesse2020reducing}
Leguesse, Y., Vella, M., Colombo, C., and Hernandez-Castro, J. (2020).
\newblock Reducing the forensic footprint with {A}ndroid accessibility attacks.
\newblock In {\em STM}, pages 22--38.

\bibitem[Li et~al., 2015]{li2015real}
Li, S., Chen, J., Spyridopoulos, T., Andriotis, P., Ludwiniak, R., and Russell,
  G. (2015).
\newblock Real-time monitoring of privacy abuses and intrusion detection in
  {A}ndroid system.
\newblock In {\em HAS}, pages 379--390.

\bibitem[Luttgens et~al., 2014]{irbook}
Luttgens, J.~T., Pepe, M., and Mandia, K. (2014).
\newblock {\em Incident Response \& Computer Forensics}.
\newblock McGraw-Hill Education Group, 3rd edition.

\bibitem[Mohammad and Alqahtani, 2019]{MOHAMMAD2019mlfortimeline}
Mohammad, R. M.~A. and Alqahtani, M. (2019).
\newblock A comparison of machine learning techniques for file system forensics
  analysis.
\newblock {\em Journal of Information Security and Applications}, 46:53--61.

\bibitem[Pagani et~al., 2019]{pagani2019introducing}
Pagani, F., Fedorov, O., and Balzarotti, D. (2019).
\newblock Introducing the temporal dimension to memory forensics.
\newblock {\em ACM TOPS}, 22(2):1--21.

\bibitem[Saltaformaggio et~al., 2015]{saltaformaggio2015vcr}
Saltaformaggio, B., Bhatia, R., Gu, Z., Zhang, X., and Xu, D. (2015).
\newblock Vcr: App-agnostic recovery of photographic evidence from android
  device memory images.
\newblock In {\em ACM SIGSAC}, pages 146--157.

\bibitem[Scrivens and Lin, 2017]{scrivens2017android}
Scrivens, N. and Lin, X. (2017).
\newblock Android digital forensics: data, extraction and analysis.
\newblock In {\em ACM}, pages 1--10.

\bibitem[Shi et~al., 2019]{shi2019jekyll}
Shi, L., Fu, J., Guo, Z., and Ming, J. (2019).
\newblock ``jekyll and {H}yde'' is risky: Shared-everything threat mitigation
  in dual-instance apps.
\newblock In {\em MOBISYS}, pages 222--235.

\bibitem[Srivastava and Tapaswi, 2015]{srivastava2015logical}
Srivastava, H. and Tapaswi, S. (2015).
\newblock Logical acquisition and analysis of data from android mobile devices.
\newblock {\em Information \& Computer Security}.

\bibitem[{Stefanko, Lukas}, 2020]{defensorid}
{Stefanko, Lukas} (2020).
\newblock {I}nsidious {A}ndroid malware gives up all malicious features but one
  to gain stealth.
\newblock
  \url{https://www.welivesecurity.com/2020/05/22/insidious-android-malware-gives-up-all-malicious-features-but-one-gain-stealth/}.
\newblock Accessed: 24.03.2021.

\bibitem[Taubmann et~al., 2018]{taubmann2018droidkex}
Taubmann, B., Alabduljaleel, O., and Reiser, H.~P. (2018).
\newblock Droid{K}ex: Fast extraction of ephemeral {TLS} keys from the memory
  of {A}ndroid apps.
\newblock {\em Digital Investigation}, 26:S67--S76.

\bibitem[{ThreatFabric}, 2020]{androidrat}
{ThreatFabric} (2020).
\newblock 2020 - year of the {RAT}.
\newblock \url{https://www.threatfabric.com/blogs/2020_year_of_the_rat.html}.
\newblock Accessed: 24.03.2021.

\bibitem[{Whittaker, Zack}, 2020]{eventbot}
{Whittaker, Zack} (2020).
\newblock {E}ventbot: {A} new mobile banking trojan is born.
\newblock
  \url{https://www.cybereason.com/blog/eventbot-a-new-mobile-banking-trojan-is-born}.
\newblock Accessed: 24.03.2021.

\bibitem[Yang et~al., 2015]{yang2015new}
Yang, S.~J., Choi, J.~H., Kim, K.~B., and Chang, T. (2015).
\newblock New acquisition method based on firmware update protocols for android
  smartphones.
\newblock {\em Digital Investigation}, 14:S68--S76.

\end{thebibliography}

\end{document}